\documentclass[sigconf]{acmart}

\usepackage{graphicx}
\usepackage{booktabs} 
\usepackage{subfigure}
\usepackage{url}
\usepackage{color}
\usepackage{listings}
\usepackage{flushend}

\copyrightyear{2020}
\acmYear{2020}
\setcopyright{none}
\acmConference{SPMA'20}{April 27, 2020}{}

\begin{document}
\title[Assessing Impact of Data Partitioning for Approximate Memory in C/C++ Code]{Assessing Impact of Data Partitioning for\\Approximate Memory in C/C++ Code}

\author{Soramichi Akiyama}
\affiliation{%
  \institution{Department of Creative Informatics, The University of Tokyo}
}
\email{akiyama@ci.i.u-tokyo.ac.jp}

\begin{abstract}
  Approximate memory is a technique to mitigate the performance gap between memory subsystems and CPUs with its reduced access latency at a cost of data integrity.
  To gain benefit from approximate memory for realistic applications, it is crucial to partition applications' data to {\it approximate data} and {\it critical data} and apply different error rates.
  However, error rates cannot be controlled in a fine-grained manner (e.g., per byte) due to fundamental limitations of how approximate memory can be realized.
  Due to this, if approximate data and critical data are interleaved in a data structure (e.g., a C struct that has a pointer and an approximatable number as its members), data partitioning may degrade the application's performance because the data structure must be split to separate memory regions that have different error rates.
  This paper is the first to conduct an analysis of realistic C/C++ code to assess the impact of this problem.
  First, we find the type of data (e.g., ``int'', ``struct point'') that is assessed by the instruction that incurs the largest number of cache misses in a benchmark,
  which we refer to as the {\it target data type}.
  Second, we qualitatively estimate if the target data type of an application has approximate data and critical data interleaved.
  To this end, we set up three criteria to analyze it because definitively distinguishing a piece of data as approximate data or critical data is infeasible since it depends on each use-case.
  We analyze 11 memory intensive benchmarks from SPEC CPU 2006 and 2 graph analytics frameworks,
  and show that the target data types of 9 benchmarks are either a C struct or a C++ class (criterion 1).
  Among them, two have a pointer and a non-pointer member together (criterion 2) and three have a floating point number and other members together (criterion 3).
\end{abstract}
\maketitle

\section{Approximate Memory Architecture}
\subsection{Overview of Approximate Memory}
Approximate main memory, or approximate memory, is a new technology to mitigate the performance gap between memory subsystems and CPUs of computers.
The main idea is to reduce the latency of main memory accesses at a cost of the data integrity (i.e., the CPU may read a slightly different data from what has been written before to the main memory) by exploiting {\it design margins} that exist in many DRAM chips today.
A design margin refers to the difference between a design parameter defined in the specification of a device and the actual value the device can be operated with.
For example, many DRAM chips can read stored data ``almost'' correctly with a few bit-flips (errors) injected to the data when some wait-time parameters are shortened than the specification~\cite{Chang2016}, resulting in access latency reduction.

Approximate memory attracts much research interest due to the ever-increasing performance gap between memory subsystems and CPUs~\cite{Hennessy2017}.
Chang~{\it et al.}~\cite{Chang2016} measure the relationship between error rates and latency reduction for a large number of commercial DRAM chips,
Das~{\it et al.}~\cite{Das2018} and Zhang~{\it et al.}~\cite{Zhang2016} prolong the interval of refreshing\footnote{A DRAM cell is a tiny capacitor and loses its charge as time goes by, thus it needs to be periodically re-charged (also known as ``refreshed'').} for strong memory cells to reduce the average latency,
and our previous work~\cite{Akiyama2019} estimates effect of approximate memory to realistic applications without simulation by counting the number of DRAM internal operations that induce errors.

\subsection{Relaxing Timing Constraints}
\begin{figure}[!t]
  \centering
  \includegraphics[width=0.75\columnwidth]{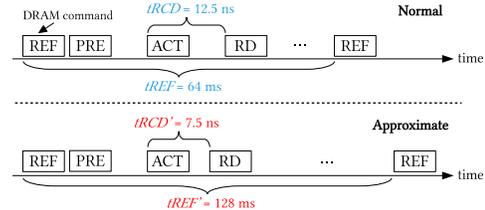}
  \caption{\label{figure:command_timeline}DRAM command sequence of normal memory (top) and approximate memory (bottom): In this example, tRCD is shortened to 7.5~ns and tREF is prolonged to 128~ms, both of which reduce the average latency.}
\end{figure}

Approximate memory on DRAM can be realized by {\it relaxing timing constraints} of DRAM chips.
Although there are other types of approximate memory such as approximate flash memory for storage that leverages multiple levels of flash programming voltages~\cite{Guo2016} and approximate SRAM based on supply voltage scaling~\cite{Chandramoorthy2019,Yang2017,Esmaeilzadeh2012}, we focus on approximate DRAM using relaxed timing constraints.
A timing constraint refers to the specification of the interval between DRAM commands issued from the memory controller,
and relaxing a constraint means either shortening or prolonging an interval (i.e., ``violating'' the specification).
Relaxing a timing constraint of DRAM reduces the access latency to main memory but may inject errors (bit-flips) to memory cells with an error rate depending on how aggressively a constraint is relaxed and other aspects such as the chip temperature~\cite{Kim2019}.

Figure~\ref{figure:command_timeline} shows an example of DRAM command sequence in normal (exact) memory and approximate memory.
It shows four representative DRAM commands: refresh (\verb|REF|), precharge (\verb|PRE|), activation (\verb|ACT|), and read (\verb|RD|).
In this example, a timing constraint called \verb|tRCD| is shortened from 12.5~ns to 7.5~ns, and one called \verb|tREF| is prolonged from 64~ms to 128~ms.
The memory controller must wait for \verb|tRCD| between a \verb|RD| command and the preceding \verb|ACT| command to ensure that the activation has finished.
Although \verb|tRCD| is defined as 12.5~ns in the DDR3-1600J specification~\cite{JEDEC_DDR3},
Chang~{\it et al.}~\cite{Chang2016} found that only a small portion of the cells experience errors even when \verb|tRCD| is shortened below it to serve the \verb|RD| command quickly.
\verb|tREF| is another timing constraint that specifies the longest interval between two \verb|REF| commands, which refresh DRAM cells to prevent them from losing stored data.
Das~{\it et al.}~\cite{Das2018} and Zhang~{\it et al.}~\cite{Zhang2016} propose to prolong this interval because many DRAM cells can retain data for more than 64~ms in practice.
Because prolonging \verb|tREF| increases the amount of time during which more useful commands are served, it reduces the average DRAM access latency.

\subsection{\label{section:approximation_granularity}Approximation Granularity}
\begin{figure}[!t]
  \centering
  \includegraphics[width=0.95\columnwidth]{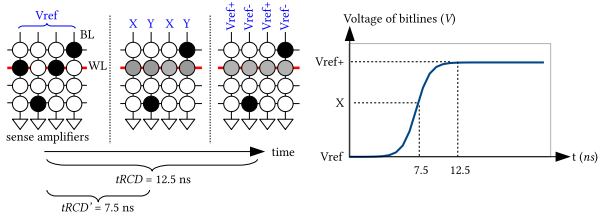}
  \caption{\label{figure:activation_timeline}An ACT command drives an entire row at the same time with a given tRCD value, forcing the minimum approximation granularity to be a row size (typically 4 KB or 8 KB).}
\end{figure}

In approximate memory that exploits design margins in the timing constraints, the granularity of approximation cannot be smaller than 4~KB (i.e., the same error rate must be applied to a continuous 4~KB or more).
This is because a DRAM command drives many memory cells in parallel to improve performance.

Figure~\ref{figure:activation_timeline} shows how an \verb|ACT| command drives multiple memory cells in parallel.
The circles show an array of memory cells, where each row has a wordline (WL) and each column has a bitline (BL).
A black cell has a value of 1 and a white cell has a value of 0, and a gray one is in an intermediate state.
An \verb|ACT| command takes the target row number as its parameter (for example, the 2$^{\rm nd}$ row in the figure).
The WL of the target row is enabled to connect the cells in the target row to the BLs.
The electric charge of the cells whose values are 1 pull up the voltages of the BLs from \verb|Vref| (the reference value) to \verb|Vref+|, which takes up to \verb|tRCD|.
Finally, the sense amplifiers sense the voltages of the BLs to fetch the values in the cells.
If \verb|tRCD| is reduced from 12.5~ns to say 7.5~ns, the sense amplifiers may fetch wrong values because the voltages of the BLs have not yet reached to \verb|Vref+| as shown in the right of the figure.
The same discussion is applicable to ``0'' cells and \verb|Vref-|.
Because a shortened \verb|tRCD| is applied to an entire row, it forces the granularity of approximation to be the size of a row, which is typically 4~KB or 8~KB.
Therefore, we cannot control the error rate with a granularity smaller than 4~KB.
Similar discussions can be applied to other timing constraints as well such as \verb|tREF| and a \verb|REF| command that reads an entire row at a time to refresh the cells in the row~\cite{Jacob2007}.

Although this work focuses only on DRAM, this limitation is also applicable to other memory technologies
because the requirement of performance is fundamental due to the ever-increasing performance gap between memory subsystems and CPUs.
For example, phase change memory (PCM) organizes memory cells as an array~\cite{Nishi2019} and injects electric pulses to an entire row at once.
If we consider realizing approximate memory with PCM, for example by reducing the length of pulses for writing, the approximation granularity is still limited by its row size.

\section{Data Partitioning and Challenge}
\subsection{Data Partitioning}
Executing an application on approximate memory requires two additional steps compared to executing it on normal memory.
First, we must identify which parts of the application's data are approximate data or critical data.
This prevents errors from being injected into critical data (e.g., pointers) and the application can yield meaningful results.
Second, we must map approximate data and critical data to different memory regions that have different error rates.
We refer to these two steps as {\it data partitioning.}

\begin{figure}[t]
  \centering
  \includegraphics[width=0.7\columnwidth]{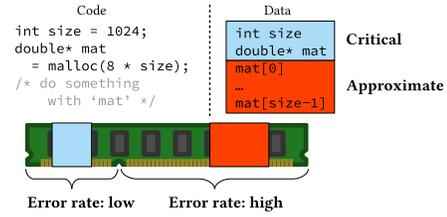}
  \caption{\label{figure:data_partitioning}Data partitioning: (1) identify critical data (size, mat) and approximate data (mat[0], ..., mat[size-1]), and
    (2) map them to memory regions that have different error rates.}
\end{figure}

Figure~\ref{figure:data_partitioning} depicts an example of data partitioning.
Assume that the application can yield acceptable results even with some errors injected to the floating point numbers stored in \verb|mat|,
then we can identify each element of \verb|mat| (\verb|mat[0]|, ..., \verb|mat[size-1]|) as approximate data and the other variables as critical data.
Note that `\verb|double* mat|' itself is critical data because it is a pointer.

After the identification, the approximate data and critical data must be mapped to memory regions with different error rates.
The size of each memory region must be at least 4~KB because the approximation granularity (the granularity at which error rates can be controlled) is 4~KB as discussed in Section~\ref{section:approximation_granularity}.

\subsection{\label{section:challenge}Challenge: Interleaved Criticality}
The challenge of data partitioning is that it can impose much overhead when approximate data and critical data are {\it interleaved} in a single data type.
The two kinds of data are defined as interleaved when they co-locate inside one element of a C \verb|struct| or a C++ \verb|class|.\footnote{The only difference of them is the access control of members in typical C/C++ implemenetations (for example, even a C struct can have member functions in gcc). Therefore, we only mention C struct hereafter without loss of generality.}
Figure~\ref{figure:example_interleaved} shows an example of interleaved approximate data and critical data.
The \verb|struct tree_node| data type has both approximate data and critical data in it,
and \verb|nodes| points to an array of \verb|struct tree_node|.
Applying data partitioning to this code requires to map the first 20 bytes (an \verb|int| and two pointers) and the last 8 bytes (a \verb|double|) of a \verb|struct node| to separate memory regions due to the large approximation granularity, splitting the region pointed to by \verb|nodes| to many pieces.
This can dramatically degrade the performance due to reduced access locality.

\begin{figure}[t]
\begin{lstlisting}
struct tree_node {
  int id;       // id of the node, critical
  struct tree_node *r; // pointer to the right child, critical
  struct tree_node *l; // pointer to the left child, critical
  double score; // score of this node, approximate
};

int size = 1000 * sizeof(struct tree_node);
struct tree_node *nodes = malloc(size);
\end{lstlisting}
\caption{\label{figure:example_interleaved}Approximate data and critical data interleaved in a single C struct: applying data partitioning to this code imposes much overhead due to reduced access locality.}
\end{figure}

Prior works~\cite{Peng2007,Curial2008,Ye2019,Jin2010} propose to convert ``an array of structures'' to ``a structure of arrays'' for improving access locality.
For example, the code in Figure~\ref{figure:example_interleaved} can be converted to have a distinct array for each member of \verb|struct tree_node|,
and then the array for \verb|score| can be mapped to a consecutive memory region with high error rate.
However, this method mitigates the data partitioning problem only when the conversion itself does not degrade access locality.
Assume that \verb|r|, \verb|l|, and \verb|score| in Figure~\ref{figure:example_interleaved} are accessed closely in time (for example, the code might change which child to traverse next depending on the score of the current node),
then splitting \verb|r|, \verb|l|, and \verb|score| into different arrays can degrade the access locality and performance.

\section{Source Code Analysis}
The research question we tackle is stated as: {\it Is data partitioning a real concern for realistic applications?}
To answer it, we analyze source code of widely used benchmarks and show that many applications potentially have approximate and critical data interleaved.

\subsection{Analysis Methodology}
Approximate memory is the most effective when an application's data that incur many cache misses are stored on it.
With this in mind, our analysis consists of two steps:
\begin{enumerate}
\item First, we find a data structure (e.g., a C \verb|struct|) accessed by an instruction that incurs the largest number of cache misses among a target application using hardware performance counters.
\item Second, we apply our criteria to qualitatively estimate the probability that the data structure found has approximate data and critical data interleaved in it.
\end{enumerate}

\begin{figure}[t]
  \centering
  \includegraphics[width=0.7\columnwidth]{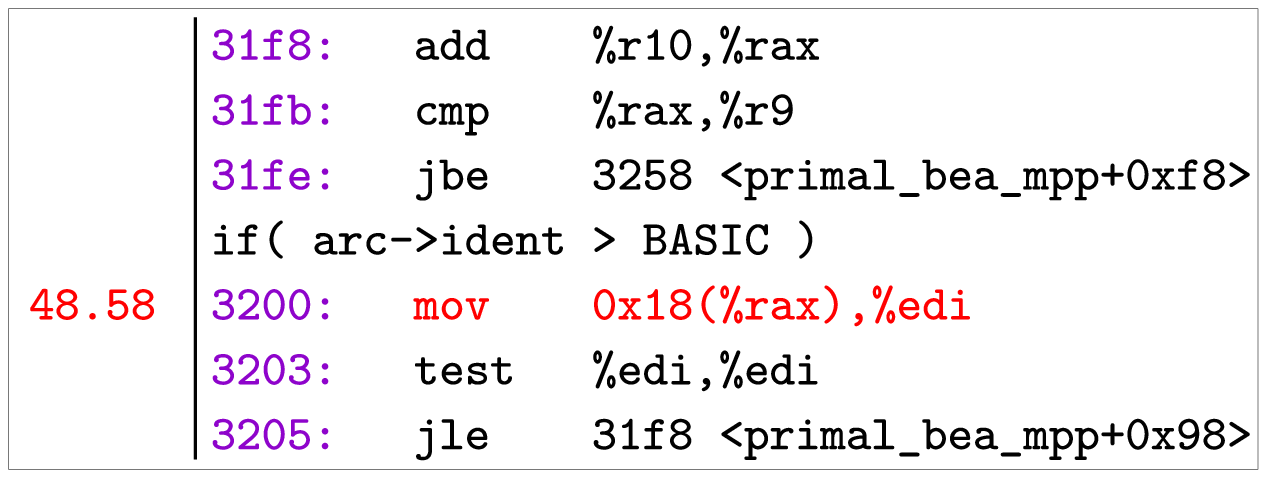}
  \caption{\label{figure:pebs_sample}Sample output of perf report: It shows an instruction, the offset of the instruction from the head of the binary, and the percentage of cache misses that it incurs (if any), from right to left. The C code, if ( arc->ident > BASIC ), corresponds to the assembly code below it.}
\end{figure}

{\bf In the first step}, we measure the number of cache misses {\bf per instruction} using Precise Event Based Sampling (PEBS) on Intel CPUs.
PEBS is an enhancement of normal performance counters that uses designated hardware for sampling to reduce the skid between the time an event (e.g., cache miss) occurs and the time it is recorded~\cite{Bakhvalov2018,Weaver2016}.
The small skid enables pinpointing which instruction in an application binary causes many cache misses.
We execute a benchmark with its sample dataset using linux \verb|perf|,
and the actual command line is `\verb|perf record -e r20D1:pp -- benchmark|'.
The parameter \verb|r20D1:pp| specifies a performance event whose {\it event number} is 0xD1 and the {\it umask value} is 0x20,
which ``{\it counts retired load instructions with at least one uop that missed in the L3 cache}'' (described in Table 19.3 of ~\cite{intel_manual}).
The parameter \verb|benchmark| is replaced with an actual command line to execute each benchmark.
Figure~\ref{figure:pebs_sample} shows a sample output of \verb|perf report|, executed after a measurement by \verb|perf record|.
The measurement is done for a benchmark called \verb|mcf| and the details of the benchmarks we analyze are described in Section~\ref{section:benchmarks}.
Each line shows, from right to left, an instruction, the offset of the instruction from the head of the binary, and the percentage of cache misses it incurs (if any).
The C code, \verb|if( arc->ident > BASIC )|, corresponds to the lines of assembly code below it.

Figure~\ref{figure:pebs_sample} shows that the instruction that incurs the largest number of cache misses in this application is \verb|mov 0x18(%rax),%edi|.
By mapping the assebmly code in Figure~\ref{figure:pebs_sample} with the C source code,
we can find that the \verb|mov| instruction accesses the \verb|ident| member of a C \verb|struct| named \verb|arc|\footnote{In fact, ident is placed in the 0x18$^{\rm th}$ byte of arc and BASIC is a compile-time constant whose value is 0. This supports our guess that the mov instruction copies ident to \%edi and the test instruction following it compares ident and BASIC (constant 0).}.
We refer to the type of \verb|arc|, \verb|struct arc|, as the {\it target data type} because it is the target of the analysis in the next step.
Please carefully note that, although this result tells that the \verb|ident| member alone incurs many cache misses,
it does not mean that we can split this \verb|struct| to exclude \verb|ident| and put an array of \verb|ident| to approximate memory.
Memory accesses to the other members of the same \verb|struct| may hit the cache only because cache misses to \verb|ident| fetch the whole part of the \verb|struct| to main memory,
in which case this splitting doubles the number of cache misses.
When the assembly code is more complex and the target data type is not obvious, we resort to human labor to find it because systematically reverse-engineering an arbitrary binary to C/C++ code is not the main focus of this work.
The same methodology is applicable to a template function as well because there is an independent piece of assembly code for each instantiation of it (i.e., no type-ambiguity exists in assembly code).

{\bf In the second step}, we analyze the target data type of each benchmark to estimate if it has approximate data and critical data interleaved.
The issue is that it is infeasible to definitively distinguish approximate data and critical data without concrete use-cases and expert knowledge of the benchmark.
There are some typical cases (e.g., pointers are {\it typically} critical data, floating point numbers are {\it typically} approximate data), but even these might be overridden by particular code or use-case.
Therefore, instead of definitively distinguishing them, we set up criteria to qualitatively predict the probability that a target data type has approximate data and critical data interleaved:

\begin{description}
\item[C1:] Is the target data type a \verb|struct| or a \verb|class|?
\item[C2:] If {\bf C1} is yes, does it include a pointer and at least one other member with non-pointer type?
\item[C3:] If {\bf C1} is yes, does it include a floating point number and at least one other member?
\end{description}

As the number of criteria applicable to a target data type increases, so does the probability that it has approximate data and critical data interleaved.
Please note the following two things. (1) If a target data type is a nested \verb|struct| (i.e., a \verb|struct| that has another \verb|struct| in it),
we ``flatten'' it so that every member becomes a non-struct type (e.g., \verb|int|, \verb|char*|) before applying the criteria.
This reproduces how a nested \verb|struct| is mapped to a memory region.
(2) We exclude member functions because they are stored in function tables located in a separate memory region from data members in typical C++ implementations, meaning that they are already partitioned and irrelevant to our analysis.
For example, if a \verb|class| has two integers and a function as its members, we consider the two integers as the only members 
and the analysis result for this data type is ``C1 = Yes, C2 = No, C3 = No''.

The intention of each criterion is as follows.
{\bf [C1]}: If the target data type is not a \verb|struct|, we can allocate the whole part of it (or an array of it) either in normal memory or approximate memory data and no data partitioning is needed.
On the other hand, if it is a \verb|struct|, it might include approximate data and critical data interleaved.
{\bf [C2]}: If the target data type includes a pointer as its member, it is highly possible that the pointer is critical data because even a single bit-flip invalidates it.
If it includes other members with non-pointer types as well, then these members might be approximate data, in which case approximate data and critical data are interleaved.
{\bf [C3]}: If the target data type includes a floating point number, it is possible that the floating point number is approximate data.
If it includes other members that are critical data, then the target data type has approximate data and critical data interleaved.
Note that {\bf C3} is not symmetric with {\bf C2} because a pointer is most probably critical data, while a floating point number can either be approximate data or critical data depending on each application and use-case.
Therefore, {\bf C2} requires a non-pointer member along with a pointer member, while {\bf C3} only requires a member with any type along with a floating point number.

\subsection{\label{section:benchmarks}Analyzed Benchmarks}
\begin{table}[t]
\centering
\caption{\label{table:benchmark}Analyzed Benchmarks: 11 from SPEC CPU 2006 (we select ones written in C/C++ and the cache miss rates are more than 20\%) and 2 graph analytics frameworks.}
\begin{tabular}{|c|c|c|}
\hline
Name & Domain & Cache Misses \\ \hline
milc & quantum simulation & 82.6\% (9.61 \%) \\
sjeng & game AI (chess) & 74.5\% (44.9 \%)\\
libquantum & quantum computing & 54.6\% (47.4 \%)\\
lbm & fluid dynamics & 49.2\% (44.5 \%)\\
omnetpp & discrete event simulation & 47.9\% (6.62 \%) \\
soplex & linear programming & 41.2\% (31.5\%) \\
gobmk & game AI (go) & 38.4\% (39.3 \%)\\
gcc & c compiler & 36.8\% (22.8 \%)\\
mcf & optimization & 33.7\% (48.6 \%)\\
dealII & finite element analysis & 33.6\% (67.1 \%)\\
namd & molecular dynamics & 21.0\% (11.3 \%) \\ \hline
Graph 500 & graph analytics (bfs) & 74.8 \% (88.8 \%)\\ \hline
GraphMat & graph analytics (PageRank) & 47.7 \% (53.5 \%) \\ 
\hline
\end{tabular}
\end{table}

\begin{table}[t]
\centering
\caption{\label{table:env}Experiment Environment}
\begin{tabular}{|c|c|}
\hline
CPU & Intel Xeon Silver 4108 (Skylake, 8 cores) \\
Memory & DDR4-2666, 96 GB (8GB $\times$ 12)\\
LLC & 11 MB (shared across all the cores) \\
OS & Debian GNU/Linux 10 \\
kernel & 4.19.0-6-amd64 \\
gcc/g++ & 8.3.0 (Debian 8.3.0-6)\\
\hline
\end{tabular}
\end{table}

Table~\ref{table:benchmark} describes the benchmarks we analyze.
Each line shows a benchmark's name, its domain, and the cache miss rate with the percentage of cache misses incurred by the instruction that incurs the largest number of cache misses.
For example, ``74.8~\% (88.8~\%)'' means that the number of cache misses divided by the number of memory accesses is 0.748,
and the number of cache misses incurred by a certain instruction divided by the number of total cache misses is 0.888.
Note that the latter can be any value between 0 and 1 independently of the former.
The values are measured in the environment in Table~\ref{table:env} and
the \verb|-O3| option is used to build the binaries.

From SPEC CPU 2006~\cite{spec2006}, we analyze 11 benchmarks whose cache miss rates are more than 20\%.
We use the largest dataset named \verb|ref| to measure the cache miss rates.
We exclude others because approximate memory is not beneficial for CPU intensive benchmarks with low cache miss rates.
We also exclude ones written in Fortran because the necessity of data partitioning is affected by the programming style, which is largely different in each programming language.
This work only focuses on C/C++ that are more often used in modern applications than Fortran.

We also analyze two graph analytics benchmarks, Graph~500~\cite{graph500} and GraphMat~\cite{Sundaram2015}.
The former is used to measure the performance of supercomputers, thus the speed is the only concern.
The latter, on the other hand, is designed to preserve programmability while maintaining the speed as much as possible.
For Graph~500, we generate the dataset with the \verb|scale| parameter set to 19 and execute breadth first search on a single core.
We use the reference implementation provided in their website~\cite{graph500}.
For GraphMat, we convert the ego-Twitter dataset of Stanford Large Network Dataset Collection~\cite{snap} into a GraphMat-compatible format and calculate PageRank on it.
The source code is taken from the github repository\footnote{https://github.com/narayanan2004/GraphMat} and we run it on a single core.

\section{Results}
\begin{table}[!t]
\centering
\caption{\label{table:results}Results of our analysis}
\begin{tabular}{|c|c|c|c|c|c|}
\hline
Name & Target Data Type &C1&C2&C3 \\ \hline
milc & struct complex &\textcolor{blue}{Y}&\textcolor{red}{N}&\textcolor{blue}{Y}\\
sjeng & struct QTType &\textcolor{blue}{Y}&\textcolor{red}{N}&\textcolor{red}{N}\\
libquantum & struct quantum\_reg\_node\_struct &\textcolor{blue}{Y}&\textcolor{red}{N}&\textcolor{blue}{Y} \\
lbm & double &\textcolor{red}{N}&-&-\\
omnetpp & class cChannel &\textcolor{blue}{Y}&\textcolor{red}{N}&\textcolor{red}{N}\\
soplex & struct Element &\textcolor{blue}{Y}&\textcolor{red}{N}&\textcolor{red}{N}\\
gobmk & Hashnode (struct hashnode\_t) &\textcolor{blue}{Y}&\textcolor{blue}{Y}&\textcolor{red}{N}\\
gcc & struct rtx\_def &\textcolor{blue}{Y}&\textcolor{red}{N}&\textcolor{red}{N}\\
mcf & arc\_t (struct arc) &\textcolor{blue}{Y}&\textcolor{blue}{Y}&\textcolor{red}{N}\\
dealII & double &\textcolor{red}{N}&-&-\\
namd & struct CompAtom &\textcolor{blue}{Y}&\textcolor{red}{N}&\textcolor{blue}{Y}\\ \hline
Graph 500 & int64\_t &\textcolor{red}{N}&-&-\\ \hline
GraphMat & float &\textcolor{red}{N}&-&-\\
\hline
\end{tabular}
\end{table}

Table~\ref{table:results} shows the analysis results.
Each row shows a benchmark, the target data type of it, and applicability of the three criteria to the benchmark (`Y' means that the criterion is applicable, and `N' means not applicable).
If {\bf C1} is `N' for a benchmark, we put `-' for {\bf C2} and {\bf C3} because these two criteria are evaluated only if {\bf C1} is `Y'.
In the target data type column, we put the original type inside parentheses if it is aliased by a \verb|typedef| declaration.
For example, ``arc\_t (struct arc)'' means that the target data type is a C \verb|struct| named arc, and it is aliased as arc\_t.
The benchmarks are executed in the environment described in Table~\ref{table:env}.
For all the benchmarks, the instruction that incurs the largest number of cache misses exists in their own code and not in any standard C/C++ libraries.

{\bf Observation 1: The target data type is a struct or a class in 9 SPEC CPU 2006 benchmarks out of 11 analyzed.}
Among them, \verb|mcf| and \verb|gobmk| have a pointer and a non-pointer member in their target data types,
and \verb|milc|, \verb|libquantum|, and \verb|namd| have a floating point number and another member in their target data types.
For the former group, if one of the members other than the pointer is approximate data, they have approximate data and critical data interleaved because the pointer is most probably critical data.
For example, our previous work~\cite{Akiyama2019} shows that \verb|mcf| can yield the same result as the error-free one even when a member of \verb|arc_t| is approximated.
For the latter group, if the floating point number is approximate data (which is the case in many applications) and another member is critical data, they have approximate data and critical data interleaved.

{\bf Observation 2: There is no benchmark whose target data type has both a pointer and a floating point number},
although this type of benchmarks (if exist) have the highest probability of having approximate data and critical data interleaved.
This might be because we exclude Fortran, which tends to be used for numerical applications.
Investigating the relationship between suitability for approximate memory and the programming language / style used in an application is a part of our future work.

{\bf Observation 3: The target data type is a non-struct type in both of the graph analytics frameworks.}
Although their source code have C \verb|struct|s that appear to have approximate data and critical data interleaved, these \verb|struct|s do not incur many cache misses.
We presume that highly optimized code for performance have less probability of including approximate data and critical data interleaved in a single data type.
Investigating the relationship between the suitability for approximate memory and the category of each application (e.g., highly optimized graph analytics) is another important aspect of future work.

\section{Discussions}
\subsection{Threats to Validity}
Our analysis methodology cannot be applied as-is when a member of a C \verb|struct| is passed to a function by reference.
For example in Figure~\ref{figure:threat}, the same function (\verb|f|) is called either by passing \verb|&s1.v| or \verb|&s2.v| as its argument.
Finding the data type that the memory region pointed to by \verb|fp| belongs to requires an investigation of stack traces and points-to analysis~\cite{Steensgaard96}.

\begin{figure}[h]
\begin{lstlisting}
struct S1 {
  double v; // probably approximate 
  double vv; // probably approximate 
} s1;

struct S2 {
  double v; // probably approximate 
  int *p; // probably critical
} s2;

void f(double *fp) {
  // do something and return the result through *fp
}

f(&s1.v); // (1): invoke f by passing s1.v by reference
f(&s2.v); // (2): invoke f by passing s2.v by reference
\end{lstlisting}
\caption{\label{figure:threat}Calling the same function by passing members of different structs by reference.
Identifying the data type that *fp belongs to requires stack traces and points-to analysis.
}
\end{figure}

Although it seems more natural for a function to take a pointer of a whole \verb|struct| such as `\verb|void g(struct S1 *sp)|',
this may appear in some cases such as when a library function returns the result through a pointer.
In our case study, the original data type was identifiable without hitting this issue in all the benchmarks.

\subsection{Analyzing Performance Implication}
This paper is focused on estimating if an application has approximate data and critical data interleaved,
but we defer quantitative analysis of how the application's performance is degraded if we really partition approximate data and critical data into separate memory regions.
One method to conduct this analysis is to actually split the target data type and measure the slowdown of the application on a real machine.
However, the issue is that although much work have been done on structure splitting, structure peeling, and structure reordering and data partitioning is essentially the same as structure splitting, there is no off-the-shelf tool to support it as far as we know.
One of the reasons is that it is very troublesome to implement these techniques so as to work correctly in general cases.
For example, old versions of \verb|gcc| support structure reordering, but the functionality was removed because it ``{\it did not always work correctly}''~\cite{gcc_4.8_release_note}.

An alternative method to qualify how an application slows down after data partitioning is to use heuristic metrics that estimate the access locality between two structure members.
Prior works on leverage these metrics to find which members should be in the same structure to maximize the performance improvement.
For example, Ye~{\it et al.}~\cite{Ye2019} introduce a metric called {\it access affinity} between two members \verb|u| and \verb|v| of a structure.
Given a memory trace, it is increamented by one if there is a pair of memory accesses to \verb|u| and \verb|v| where the number of memory accesses to other members in-between is less than a threashold.
Because this metric captures access locality of two members of a structure, it can also be used to estimate how harmful to split two members into separate memory regions.
Other studies such as~\cite{Miucin2018,Jin2010} propose similar metrics.
Applying these metrics is easier than implementing data partitioning because they are based on memory traces of the original program.

\section{Related Work}
To the best of our knowledge, this work is the first to investigate the data partitioning problem for approximate memory.
We discuss related work in the context of error rate controlling.

Nguyen~{\it et al.}~\cite{Nguyen2020} propose a method that partially mitigates the data partitioning problem.
It transposes rows and columns of data layout inside DRAM so that a chunk of data is stored across many rows that have different error rates.
This enables protection of important bits (e.g., the sign bit of a floating point number) while aggressively approximating less important bits.
This mechanism is effective for DNNs because they require the whole part of a large weight matrix at once
and the number of memory accesses do not increase regardless of the data layout.
However, it is not effective in general cases where memory is accessed with smaller granularity.

Mapping data into memory regions with different error rates depending on its criticality is commonly proposed.
Liu~{\it et al.}~\cite{Liu2011} partition a DRAM bank into bins with proper refresh interval and ones with prolonged refresh interval.
Each data is store into either type of bins depending on the criticality specified by the programmer.
Although they do not discuss the minimum bin size, it cannot be smaller than a DRAM row (typically 4~KB or 8~KB) as we discuss in Section~\ref{section:approximation_granularity}.
Chen~{\it et al.}~\cite{Chen2016} propose a memory controller that maps data into different DRAM banks with different error rates depending on the criticality of the data.
Because this method is bank-based, the approximation granularity is limited to the bank size.
A typical DDR3/DDR4 DIMM module has 2~GB to 16~GB with either 8 or 16 banks, resulting in a typical bank size of 256~MB to 2~GB.
Raha~{\it et al.}~\cite{Raha2017} advance a previous work~\cite{Liu2011} by measuring each bin's error rate at a given prolonged refresh interval
and assigning them to approximate data in the ascending order of the error rate.
They realize the bin size (or ``page size'' in their terminology) of 1~KB by measuring the average error rate per 1~KB.
Although this approach could be further pursued to realize smaller page sizes,
it still cannot control error rates per byte as it just measures them to use proper pages for given criticality.

Applying lower supply voltages to store approximate data is often done especially with SRAM~\cite{Chandramoorthy2019,Yang2017,Esmaeilzadeh2012}.
Esmaeilzadeh~{\it et al.}~\cite{Esmaeilzadeh2012} propose a dual-voltage SRAM architecture to implement approximate data types proposed by Sampson~{\it et al.}~\cite{Sampson2011}.
This architecture also suffers from the same problem we discuss in this paper because it changes the supply voltage at the row granularity of an SRAM subarray,
although a row of a subarray is smaller than a row of the entire SRAM.

Tovletoglou~\cite{Tovletoglou2020}~{\it et al.} propose an end-to-end framework that ensures the availability of VMs hosted on approximate memory.
The framework provides an OS support and its APIs to allocate approximate data to memory control units operated by reduced refresh rate, while maintaining the ability to exploit memory-level parallelism.
Because the framework leverages the existing memory allocators implemented in Linux, the minimum size of a critical / approximate memory region is 4~KB,
making it suffer from the same problem we discuss in this paper.

\section{Conclusions and Future Work}
Approximate memory is a new technique to reduce the main memory access latency,
but has a potential problem in data partitioning when approximate data and critical data are interleaved in applications' data structures.
We are the first to assess the impact of this problem by means of analyzing realistic C/C++ code.
We introduce three criteria to qualitatively estimate if the data type that incurs the largest number of cache misses in an application has approximate data and critical data interleaved,
and applied them to 11 SPEC CPU 2006 benchmarks and 2 graph analytics frameworks.
As a result, we found that the data type that incurs the largest number of cache misses are either a C \verb|struct| or a C++ \verb|class| in 9 benchmarks.
Among them, two have a pointer (possibly critical) and a non-pointer member interleaved
and three have a floating point number (possibly approximate) and other members interleaved.

Future work includes two directions: (1) categorizing workloads by their domains, programming language / styles used and other aspects,
and (2) quantitatively analyzing the performance implication of data partitioning to each workload.
For the second part, no off-the-shelf tool that enables data partitioning is available even though much work have been done on structure splitting,
mostly because implementing structure splitting for general cases is very difficult.
Instead, we can use existing metrics~\cite{Ye2019,Miucin2018,Jin2010} that predict the benefit of structure splitting for a given structure.

\begin{acks}
  This work was supported by JST, ACT-I Grant Number JPMJPR18U1, Japan.
  The author thanks the anonymous reviewers for their valuable feedback.
\end{acks}

\bibliographystyle{ACM-Reference-Format}
\bibliography{SPMA2020}

\end{document}